\documentclass[12pt,preprint]{aastex}

\slugcomment{Accepted for publication in AJ}

\shorttitle{Optical Imaging and Photometry of VLIRGs}
\shortauthors{Arribas et al.}

\begin{document}


\title{Optical Imaging of Very Luminous Infrared Galaxy Systems: Photometric Properties and Late Evolution}


\author{Santiago Arribas\altaffilmark{1,2}, Howard Bushouse, and Ray A. Lucas}
\affil{Space Telescope Science Institute, 3700 San Martin Drive,
 Baltimore, MD 21218, USA}
\email{arribas@stsci.edu, bushouse@stsci.edu, lucas@stsci.edu}

\author{Luis Colina}
\affil{Instituto de Estructura de la Materia, CSIC, Serrano 119, 28006-Madrid, Spain}
\email{colina@damir.iem.csic.es}

\and

\author{Kirk D. Borne}
\affil{George Mason University, School of Computational Sciences, NASA Goddard Space Flight Center, Greenbelt, MD  20771, USA  }
\email{borne@mail630.gsfc.nasa.gov}


\altaffiltext{1}{Affiliated  with the Space Telescope Division of the European Space Agency, ESTEC, Noordwijk, Netherlands.}
\altaffiltext{2}{On leave from the Instituto de Astrof\'\i sica de Canarias (IAC) and from the Consejo Superior de Investigaciones Cientificas (CSIC), Spain.}


\begin{abstract}

A sample of 19 low redshift (0.03$<$z$<$0.07) very luminous infrared galaxy (VLIRG: $10^{11}L_\odot< $ L[8-1000 $\mu$m] $ < 10^{12} L_\odot$) systems (30 galaxies) has been imaged in $B$, $V$, and $I$ using ALFOSC with the Nordic Optical Telescope. These objects cover a luminosity range that is key to linking the most luminous infrared galaxies with the population of galaxies at large.  As previous morphological studies have reported, most of these objects exhibit features similar to those found in ultraluminous infrared galaxies (ULIRGs), which suggests that they are also undergoing  strong interactions or mergers. We have obtained photometry for all of these VLIRG systems, the individual galaxies (when detached), and their nuclei, and the relative behavior of these classes has been studied in optical color-magnitude diagrams.  The observed colors and magnitudes for  both the systems and the nuclei lie parallel to the reddening vector, with  most of the nuclei having redder colors than the galaxy disks.  Typically, the nuclei comprise 10 percent of the total flux of the system in $B$, and 13 percent in $I$. The photometric properties of the sample are also compared with previously studied samples of ULIRGs. The mean observed optical colors and magnitudes agree well with those of cool ULIRGs. The properties of the nuclei also agree with those of warm ULIRGs, though the latter show a much larger scatter in both luminosity and color. Therefore, the mean observed photometric properties of VLIRG and ULIRG samples, considered as a whole, are indistinguishable at optical wavelengths. This suggests that not only ULIRG, but also the more numerous population of VLIRGs, have similar rest-frame optical photometric properties as the submillimeter galaxies (SMG), reinforcing the connection between low-{\it z} LIRGs -- high-{\it z} SMGs. When the nuclei of the {\it young} and {\it old} interacting systems (classified according to a scheme based on morphological features) are considered separately, some differences between the VLIRG and the ULIRG samples are found. In particular, although the young VLIRGs and ULIRGs seem to share similar properties, the old VLIRGs are less luminous and redder than old ULIRG systems. If confirmed with larger samples, this behavior suggests that the late-stage evolution is different for VLIRGs and ULIRGs. Specifically, as suggested from spectroscopic data, the present photometric observations support the idea that the activity during the late phases of VLIRG evolution is dominated by starbursts, while a higher proportion of ULIRGs could evolve into a QSO type of object. 

\end{abstract}




\section{Introduction}

Luminous infrared galaxies have been the subject of numerous studies
over the past years (see, for instance, Sanders \& Mirabel 1996;
Veilleux et al. 2002, and references therein). Many of these works,
both from the ground and space-based, have been focussed on the most
energetic objects: the Ultraluminous Infrared Galaxies (ULIRGs: $L_{ir}=L[8-1000\mu m]  > 10^{12} L_\odot$) (e.g. Sanders et al. 1988; Melnick and Mirabel, 1990; Leech et al. 1994; Murphy et al. 1996; Clements et al.
1996; Surace et al. 1998 and 2000; Borne et al. 2000; Colina et al. 2001; Farrah et al. 2001; Kim et al. 2002, Bushouse et al. 2002 and references
therein). These studies have found that the vast majority of ULIRGs are
strongly interacting or advanced merger systems. The high IR luminosities are attributed to dust
emission, with the heating source being varying combinations of
interaction-induced starbursts and active galactic nuclei (AGN). ULIRGs also appear to be
forming moderately massive ({\it L*}) field ellipticals (e.g. Genzel et
al. 2001 and references therein).

At a lower energy, the Very Luminous Infrared Galaxies
(VLIRGs: $10^{11}L_\odot<  L_{ir} < 10^{12} L_\odot$, see Section 2.1) have
not been the subject of as much scrutiny (although see, for instance,
Lawrence et al. 1989; Kim et al. 1995; Wu et al. 1998 a and b).
However, this subclass of lower
luminosity objects has become increasingly interesting for several reasons. First, the above mentioned
studies on ULIRGs have shown that many of the fundamental properties of these objects,
such as the frequency of interactions, the interaction phase, the
frequency of AGN, etc., all appear to correlate with the IR luminosity
(Sanders \& Mirabel 1996, and references therein).  VLIRG studies offer the possibility to
analyze these properties over a wider luminosity range. Also, VLIRGs represent a much larger fraction of galaxies in the local
universe than ULIRGs. In particular, the local density of VLIRGs is
$\sim$2 orders of magnitude higher than the density of ULIRGs (see, for
example, Soifer {\it et al}. 1987; Saunders {\it et al}.  1990).
Moreover, the infrared luminosities of VLIRGs are between those of the
ULIRGs and local field spirals, for which typically $L_{ir} =
10^{10}-10^{11} L_\odot$ (Rieke and Lebofsky, 1986). Therefore, VLIRGs
represent a key link between the ULIRGs and the population of galaxies
at large. VLIRGs are also
believed to be low-redshift analogs of the galaxies that give rise to the
far-IR background (Hauser et al., 1998) (e.g. high-redshift sub-mm galaxies, SMGs or
SCUBA sources;  Smail et al. 1998). The nature of the SMGs is not yet clear. Some of them could be ULIRGs, as has recently been suggested by Frayer et al. (2003) from their near infrared (rest frame optical) magnitudes and colors. However, the majority could be related to the significantly more numerous class of VLIRGs. Therefore, well studied local samples of VLIRGs 
are valuable when establishing the relationship with the high redshift populations. This is especially true in light of the ongoing deep surveys such as, for instance, the Great Observatories Origin Deep Survey (GOODS, Dickinson et al. 2003; Giavalisco et al. 2004), and the Ultra Deep Field (Beckwith et al. 2003).   

This paper presents recently obtained ground-based optical $B$,
$V$, and $I$ images for a sample of VLIRGs. In Section 2, we comment on previous optical imaging surveys of VLIRGs,
describe the characteristics of our sample, and give details about the observations. Section 3 describes the data reduction process. In Section 4, the photometric properties are presented and compared with previously studied  samples of ULIRGs. 
In Section 5, we present our conclusions. An appendix gives additional comments on 
individual objects.

\section{Sample and Observations}

\subsection {Previous optical imaging VLIRG studies: the properties of the sample}

Previous optical imaging studies of luminous infrared galaxies have included
some VLIRGs. When reviewing previous works  it is important to establish a common definition of
VLIRGs.  As mentioned in Section 1, we will consider VLIRGs as those galaxy systems
having $10^{11}L_\odot< L_{ir}=L[8-1000\mu m] < 10^{12} L_\odot$, for $H_o$= 75 kms$^{-1}Mpc^{-1}$.
Sanders and Mirabel (1996) give a prescription to compute $ L[8-1000\mu m] $ based on the four $IRAS$ bands. However, many of these objects
have no detections in the IRAS 12 $\mu$m and 25 $\mu$m bands, and
therefore it is not possible to derive IR luminosities in that
manner. However, for the type of galaxies in this sample, the
luminosity inferred from the 60$\mu$m band is roughly half the luminosity in
the range [8-1000$\mu$m]  (see Bushouse et al. 2002 and references
therein). Therefore, if the luminosity from the 60$\mu$m-band is used
($L_{60}$) VLIRGs should lie in the range $10^{10.7}L_\odot< L_{60} < 10^{11.7} L_\odot$. 
Also note that the derived luminosities are
sensitive to the $H_o$ value
considered. For definition purposes, and all through this paper we will use $H_o$= 75 kms$^{-1}Mpc^{-1}$.

Taking all this into account, Lawrence et al. (1989) imaged $\sim$ 30
VLIRGs in the $R$-band out of their sample of 60 objects. No photometric values were reported, since this sample was
used to perform a morphological analysis on the percentage of
luminous galaxies that are disturbed or in interacting systems. Leech
et al. (1994) also included  7 VLIRGs in their $R$-band imaging survey. Again in this case no photometry was reported.
Further optical imaging of VLIRGs includes that by Murphy et al (1996, 5 objects in the
$r-Gunn$), and Melnick and Mirabel (1990; 1 VLIRG  imaged in $R$).
Clements et al. (1996) obtained optical imaging ($R$-band) of 60 objects but, as claimed by these authors,
the sample contains basically ULIRGs, not VLIRGs
(note that they used $H_o$ = 100  kms$^{-1}Mpc^{-1}$). In summary, most of the previous optical imaging studies of luminous
infrared galaxies are related to the most luminous objects (i.e.
ULIRGs). They were focussed on inferring morphological properties (e.g.
interaction, merger rates) based on single filter imaging (generally
$R$), and no photometric or color values were reported.

Here, we present multi-wavelength optical imaging and photometry for a sample of 19 low
redshift ($<z> = 0.047$, 0.03$<$z$<$0.07) VLIRG systems (30 galaxies). The objects
were mainly selected from the 1Jy and 2Jy redshift samples of Straus et al. (1992) and
Fisher et al. (1995), respectively, with a few cases from other sources (see Table 1). 
From all potential candidates, those suitable for
observations from La Palma (latitude $\sim$ +28 degrees) during the
summer were finally selected. The current sample is not complete,
but it likely covers the full range of properties of these objects. Eight of these objects have only upper limits for their $f_{25}$ and/or $f_{60}$ IRAS fluxes, and therefore for them it was not possible to compute their integrated [8-1000$\mu$] infrared luminosity following the prescription given by 
Sanders and Mirabel (1996) (i.e. only upper limits). Twelve systems in the sample can be classified as cool (i.e. $f_{25}/f_{60} <$ 0.2), two as warm, and five are undefined due to the fact that only an upper limit is available for $f_{25}$. Six objects have optical spectroscopy by Veilleux et al. (1995) and/or Wu et al. (1998b). Further details about the objects in the sample can be found in Table 1. 

\subsection {Observations}

We have used the Nordic Optical Telescope (NOT) in combination with
ALFOSC (Andalucia Faint Object Spectrograph and Camera) to observe our
sample in imaging mode on the nights of 2002 July 08 - 2002 July 11. ALFOSC is a 2048 x 2048 pixel Loral/Lesser CCD with a pixel size of 15
microns and a plate scale of 0.188$''$ per pixel. Ten of the 19
galaxies, primarily those exhibiting detailed, extended structure on
the Digitized Sky Survey images, were imaged in 3 filters, B74 (440nm),
V75 (530nm), and I12 (797nm), and the remaining  were imaged in 2
filters, B74 and I12. Observations made use of a small 4-point dither
done in a skewed diagonal pattern so as not to shift along the same CCD
rows and columns.  These small dithers were made in order to offset the
effects of any chip defects or problem areas such as bad or hot pixels
during the later image combination for each galaxy. The seeing was under 1 $''$ during most of the time (see Table 2).
Calibration exposures of a field in the globular cluster M92 were taken in the same
manner on the nights of 2002 July 08-10. Flat fields were obtained by
exposure on a blank field in twilight.  The calibration of the data is
discussed in more detail in Section 3.  Table 2 gives the details of
total exposure times, filters, and observation dates for each galaxy.

\subsection{Comparison ULIRG Samples}

Throughout this paper we will compare the present sample of VLIRGs with previously studied samples of ULIRGs.
Those come from the works by Surace et al. (1998) and Surace, Sanders, and Evans (2000). Surace et al (1998) studied a sample of 9 warm ULIRGs (w-ULIRG, if $f_{25}/f_{60} >$ 0.2), for which they obtained HST/WFPC2 $B$ and $I$ imaging. They also obtained aperture photometry for the nuclei, as well as for the knots of star formation in the Johnson system. They did not use a standard size aperture, but rather adjusted its size (and sometimes its shape) to the characteristics of the nuclei and knots (typical circular aperture ranged between 0.4 and 0.8 arcsec in diameter). The mean redshift for their sample is 0.094 $\pm$ 0.043. Surace, Sanders, and Evans (2000) presented ground-based $B$ and $I$ imaging for a sample of 14 cool ULIRGs (c-ULIRG, if $f_{25}/f_{60} <$ 0.2). They used a fast tip/tilt image stabilizer resulting in a FWHM of $\sim$ 0.3 $''$ for point sources. In this case, nuclear (in an aperture of  2.5 kpc in diameter) as well as global photometry for the complete system was provided. This sample of cool ULIRGs has a mean redshift of 0.087 $\pm$ 0.042. Note that the VLIRG, cool-ULIRG, and warm-ULIRG samples have different linear resolutions. Specifically, the VLIRG sample, which was observed under an average seeing of 0.9$''$ and has a $<z> \sim$ 0.047, has a linear resolution lower than the cool-ULIRG and warm-ULIRG samples by factors of $\sim$ 1.5 and 4, respectively.  This fact should not affect the comparison between the nuclear photometry of the VLIRGs and the cool-ULIRGs. In these two cases the photometry was carried out using an aperture of 2.5 kpc in diameter, which is equivalent to an angular aperture of 2.8 arcsec at the mean distance of the objects in the VLIRG sample (i.e. a factor three larger than our mean seeing). For the warm-ULIRG sample, the different linear resolutions may introduce some uncertainty, though this is estimated to be small, taking into account the intrinsic scatter of this sample (see Section 4.3).  In principle, the different linear resolution may also bias the interaction classification in the sense that objects in samples observed with lower linear resolution tend to be classified as single-nucleus objects and, therefore, as belonging to later interaction classes. This possibility will be discussed for our particular sample in Section 4.1.     

\section{Data Reducion}

\subsection {Individual Images}

Initial data processing and reduction was as follows. All {\sc bias} exposures 
for a given night were divided by the first one of the night in order to check 
for variations, and the resulting statistics were used to identify {\sc bias} exposures with
anomalous levels which were then excluded from further processing. {\sc Superbias}
images were then created on a per night basis by combining the remaining
good {\sc bias} exposures in roughly equal numbers from the beginning and end of each
night, randomly excluding some of the {\sc bias} exposures when there was an excess 
at either the beginning or end of a given night. The {\sc bias} was then subtracted
from each image on a per night basis using the {\sc superbias} image for each night, 
and the images were trimmed to a size of 2001 x 2001 pixels. Next, {\sc flat} images
were examined for stars and other obvious undesirable features and those with
such characteristics and, for example, unusually longer exposure times, were
excluded from the subsequent combination of the {\sc flats}. {\sc Superflat} images were
made for each filter by combining all the remaining good {\sc flats} on a per filter basis across the 4 nights. The images were weighted  
by the mean of the counts during this combination. These mean count statistics
were then used to normalize all the {\sc flats} to a value of 1, and then the science
and calibration images were flat-fielded using the resulting {\sc superflats}.

Observations obtained using the  $I$-band filter suffer from fringing due
to night sky emission lines. A fringe calibration image was created by
median-combining 52  $I$-band science images taken from all nights of
observations. An iterative procedure was then used to determine an
optimal scale factor for the fringe image which, when subtracted from
each $I$-band science image, minimized the residuals in the sky background.

\subsection {Combined Images}

The fully calibrated and corrected science images for each target were then
registered and combined. The IRAF task {\sc imalign} was used to compute image
offsets within each dither pattern and to then shift each of the individual
images to a common reference. Six stars common to each frame were used to
compute the offsets and third-order spline interpolation was used to
produce the shifted images. All images for a given target were shifted to
the reference frame of the $B$-band observations. The IRAF task {\sc imcombine}
was then used to combine the shifted images for each filter set, using
an iterative clipping technique to reject pixels affected by cosmic rays.

\subsection {Photometric Calibration}

For calibration purposes we observed the
globular cluster M92 (NGC6341) during the nights July 8 through July10. This cluster is included in the
list of photometric standard fields by Stetson (see Stetson and
William, 1988; and http://cadcwww.hia.nrc.ca/standards/), and it has
been used at the NOT for a program to monitor the stability of the
photometric zero points with ALFOSC.

We selected 5 stars from this catalogue (D-5, D-6, D-7, D-9,
D-23), which were located in the central parts of the chip (i.e. where
the program objects were observed). The instrumental magnitudes were
measured using the task {\sc phot} in IRAF. For the
'instrumental to photometric' transformations, we assumed the mean
color and extinction coefficients obtained from the
aforementioned monitor program. The standard deviation of the 15 zero
point values per filter inferred from each star/night combination were
on the order of 0.01 mag, without a discernible trend with night. The
mean inferred zero points (25.735, 26.680, and 24.892 for $B$,$V$, and $I$,
respectively) were also in good agreement with those obtained in the
monitor program.

For the night of July 11, we did not have M92 observations. However, during that
night photometric conditions were as good as in the previous nights.
This was also checked with the Carlberg Automatic Circle Catalogue (at
http://www.ast.cam.ac.uk/$\sim$dwe/SRF/camc\_extinction.html ), which
provides measurements of the extinction at the observatory every
night.  The very good instrumental stability found during the three first
nights suggests that  the same transformations can be safely applied
for the fourth. In addition, there was one object (IRAS 2124+2342) which
was observed on July 10th and 11th and, when applying the same
transformations to  both nights, the photometric magnitudes agree to within
0.01 - 0.02 mag in all common filters.

The equations to transform counts in the images to magnitudes are:\\
$B = -2.5 log (counts/s) + 25.735 + 0.06\times(B-V) - 0.292\times{\it am}$,\\
$V = -2.5 log (counts/s) + 25.680  - 0.06\times(V-I) - 0.179\times{\it am}$,\\
$I  =  -2.5 log (counts/s) + 24.890  - 0.05\times(V-I) - 0.080\times{\it am}$,\\

where {\it am} is the air mass. The absolute magnitudes were derived in the same manner as in Surace et al. (1998). 

\section {Results}

\subsection {Morphology: Interaction Class}

In Figure 1, we present color images for all objects.
For those with only $B$ and $I$  frames we used the $I$ image for both red and green inputs, and the $B$ for blue. For
cases  where there are 2 rather widely separated and/or large galaxies, we split them into 2 frames, one for each of the
galaxies, so the details can be better seen.
 
In Figure 2 we present the individual $B$, $V$, and $I$ reduced images in a grayscale format. The $B$ image 
is scaled with two different limits in order to highlight the faint large scale structures. This filter was used for the deep stretches, since the faint  extended emission in most of the systems shows up the strongest in $B$ (see Section 4.3).  
 
We have classified  each object using the scheme first proposed by Surace (1998) and described also by  Veilleux et al (2002). The main characteristics of each class are (see further details in Veilleux et al. 2002): 
 Class I: First approach. Early stage of the interaction.  In this phase the galaxy disks remain relatively unperturbed, and there are no signs of tidal tails or bridges. 
 Class II: First contact. The disks overlap, but strong bars or tidal tails have not been formed yet.
 Class III: Pre-merger. Two identifiable nuclei, with well developed tidal tails and bridges. Class IV: Merger. Prominent tidal features, but only one nucleus. Class V: Old Merger. Disturbed central morphologies, but no clear signs of tidal tails. 

Four of the authors independently classified each of the objects in our sample. In general there was good agreement. The adopted classification is included in Table 3, which shows that one VLIRG was classified as I, one as II, nine as III, none as IV, and eight as V.  Taking into account this distribution we found that if the different classes are grouped into  two (i.e. {\it young} = I + II + III and  {\it old} = IV + V), the uncertainties in the classification were substantially reduced. In fact most of the doubts were in distinguishing classes I and II, II and III, or IV and V,  but there were virtually no doubts about whether the object should be classified as III or IV (only for IRAS 20550+1656 where one of us selected IV versus the other three who gave III as first option). 
However, in four cases, doubts about whether the system should  be classified as very young (I-II) or very old (V) arose, depending on the assumptions about which galaxies in the field of view actually form the system. To some degree that was the case for IRAS 16396+7814, IRAS 17487+5637, IRAS 19171+4707, and IRAS 21248+2342, though in all these cases there was a majority opinion (i.e. the values adopted in Table 3). The classification for the objects in the ULIRG samples was already provided by Veilleux et al. (2002), except in a few non-problematic cases.  The distribution for the cool ULIRGs was: nine as III, four as IV, and one as V. For the warm ULIRGs, three were classified as III, five as IV, and one as V. As mentioned before, effects of distance and/or spatial resolution may, in principle, introduce some bias in the classification, with a general trend of classifying as 'single-nucleus objects' those  observed with lower, poorer linear resolution. Since our VLIRG sample has lower linear resolution than the ULIRG samples, we could have classified as classes IV (or even V) objects otherwise possibly classified as III. Although we cannot rule out this possibility, for the present sample this is unlikely since we ended up with no objects classified as IV. Furthermore, as mentioned above, for the analysis performed in the current paper, we have grouped the objects in only two categories, which will minimize this type of inhomogeneity among samples. In Appendix A, morphological details for each individual system are also included.

\subsection {Photometry} 

The photometric measurements for the program objects were done using the tasks {\sc phot}, and {\sc polyphot} in IRAF. 
The task {\sc phot} was used to measure the magnitudes through a circular aperture of 2.5 kpc in diameter at the distance of
the galaxy, and {\sc polyphot} for obtaining magnitudes of individual
galaxies within pairs with irregular morphologies. The apertures for the galaxies and systems reached the sky level (see below). 

The mean sky background was measured from a nearby $clean$ area and
subtracted accordingly. The sky had  mean values of 
0.54 $\pm$ 0.1 ($B$), 1.18 $\pm$ 0.3 ($V$), and 3.8 $\pm$ 1.2 ($I$) counts/sec/pixel. This is equivalent to 22.9 $\pm$ 0.2 ($B$), 21.8 $\pm$ 0.25 ($V$), and 19.7 $\pm$ 0.25 ($I$) mag/$''$, just applying equations shown in Section 3.3. The instrumental magnitudes were transformed
to photometric magnitudes in the Stetson system, using the equations
given in Section 3. For the objects without $V$ magnitude and therefore without $(B-V)$ and
$(V-I)$ colors,  the mean values for the rest of the sample were used to compute the corresponding magnitude using equation (1). The
standard deviation of these mean values is relatively small (i.e. $\sim$ 0.2
mag), introducing an uncertainty in the final magnitude of 0.01 mag (1
sigma). Specifically, we assumed  $<B-V>$= 0.8  and $ <V-I>$ = 1.2 .

Table 3 gives the photometry for the program objects. In Table 4, the mean absolute magnitudes, colors, and redshifts for different groups are presented. There, we can see that the mean observed $M_B$ and $M_I$ for the VLIRG systems (-20.48$\pm$0.67 and  -22.44$\pm$0.45 ) agree well with the corresponding values for the cool ULIRGs (-20.58$\pm$0.77 and -22.26$\pm$0.57). There is a similar agreement between the nuclei of the VLIRG and ULIRG samples (-17.85$\pm$0.80 vs -18.04$\pm$1.68, and -20.10$\pm$0.45 vs -20.11$\pm$1.35 for $M_B$ and $M_I$, respectively). The relatively large scatter of the ULIRG sample is mainly due to the warm subsample. In any case, the mean magnitude and color values for the VLIRGs also agree with the subsample of cool ULIRGs, which has a substantially lower scatter. Table 4 also shows the relative luminosity of the nuclei with respect to the whole system. The fact that for all the samples, the images show a relatively higher flux concentration in $I$ than in $B$ suggests that obscuration is higher in the inner regions. This is also confirmed by an even higher concentration ($\sim$ 0.4) at near infrared wavelengths ($H$-band), as has been reported by Colina et al. (2001). However, we point out that if the rms of individual values is taken into account, the present data are compatible with equal concentration in the $B$ and $I$ images.  Table 4 also suggests similar concentrations for the different interaction classes in the $B$-band. In the $I$-band, there seems to be a trend in the sense that late interaction classes have a more concentrated emission. The lower scatter for the late classes indicates that the ratio $L_n/L_s$ is better defined for these systems. The comparison between the VLIRG and ULIRG samples also suggests that ULIRGs have a relatively more powerful nucleus than the VLIRGs by a factor of about 1.3. Specifically, on average a VLIRG nucleus carries about 10 percent  of the total flux of the system in $B$, while a ULIRG nucleus carries about 13 percent. In $I$, these figures are 12.7 and 17.3 percent, respectively. However, again in this case, if the scatter is taken into account, the present results are compatible with similar concentration for the VLIRGs and the ULIRGs. In the next section we will discuss in more detail the behavior of the different groups.    

\subsection {Color-Magnitude diagrams}

\subsubsection {Optical properties of VLIRGs: Systems and nuclei}

In Figure 3, we present the $M_B -(B-I)$ color-magnitude diagram for the objects in the sample. In this figure we have distinguished between the nuclei (defined within a 2.5 kpc aperture), and the whole systems (i.e. including all individual galaxy fluxes, when more than one galaxy forms the system). Lines connect nuclei and systems of the same object. This figure shows that both the systems and the nuclei have well defined distributions parallel with each other. Typically, the systems are about 2.5 mag brighter than the nuclei in $B$ ($\sim$ 2.3 mag in $I$). The two warm VLIRGs in our sample (IRAS 20135-0857, IRAS 20210+1121) do not show any special location in the diagram. The range  in both color and absolute magnitude is lower for the systems than for the nuclei. In fact, the systems expand from $B-I$= 1.2 to 2.8, and from $M_B$= -19 to -22,  while for the nuclei these range, respectively, from 0.8 to 3.1 and from -16.5 to - 20. The higher compactness of the system sequence is due to the fact that, while most of the systems are somewhat bluer than the nuclei, some exceptions to this behavior occur for the bluest systems. In Figure 1 we can also appreciate this general trend, according to which the disks and extended features are, in most of the cases, relatively blue compared with the innermost regions. The cases for which the inner regions are bluer than the outer parts likely represent nuclei with large starburst and/or AGNs. The larger scatter along the nuclei sequence may also indicate that nuclei have a large variety of reddening/extinction conditions.  

Interestingly, the nuclei and systems distributions run parallel to the reddening vector. This also suggests that the intrinsic, reddening corrected, magnitudes and colors may define more compact distributions. Unfortunately, but for a few exceptions, there are no spectroscopic measurements from which the reddening correction can be worked out for each individual system/nucleus. The exceptions are the three systems (6 nuclei) observed by Veilleux et al (1995): IRAS 16104+5235, IRAS 16577+5900, and IRAS 18329 + 5950. These six nuclei give a mean E($B-V$) of 1.18 $\pm$ 0.6 from the H$\alpha$/H$\beta$ ratio. This could be a representative mean extinction value for the nuclei in the sample, according to the E($B-V$) distributions found by Veilleux et al. (1995) from 114 objects from the IRAS Bright Galaxy Survey (median E($B-V$) = 1.13), and 86 from the IRAS Warm Galaxy Survey (median E($B-V$) = 0.91). Using the Cardelli, Clayton, and Mathis (1989) extinction law for R$_v$=3.1, the mean reddening corrections for these nuclei would be 1.8 and 4.9 magnitudes  in $M_I$ and  $M_B$, respectively. Under these assumptions the mean intrinsic magnitude and color for the nuclei in the VLIRG sample would be: $< M_B>_o$ = -22.75, and $(B-I)_o$ = - 0.85. Although the absolute magnitude is in rather good agreement with that of the QSO Mrk 1014 which is considered a 'zero point' in the color-magnitude diagram for 'infrared-loud' classical QSOs (Surace et al 1998), the corrected color, $<B-I>_o$, is more than one magnitude bluer than that of Mrk 1014. If a lower reddening correction is assumed to fit the Mrk 1014 color, $< M_B>_o$ would be about 1.5 magnitudes fainter than the one for this quasar. Therefore, the nuclei of the VLIRGs cannot be fit to Mrk 1014 in this color-magnitude diagram by changing the mean assumed reddening. 

Alternatively, the present nuclear data can also be compared with the predictions of the starburst models. 
In Figure 3, we have represented the temporal sequences generated with STARBURST99 (Leitherer 
et al. 1999), for an instantaneous burst of 10$^9 M_\odot$, and a continuous starburst with a 
rate of 50$M_\odot$/yr. A Salpeter IMF, with masses between 0.1 and 120 $M_\odot$, and solar 
metallicity were assumed in both cases. If the nuclear data are interpreted under the assumption of a continuous burst, several ages and rates of star formation are possible. Both young bursts ($\sim$ $10^6$ yr) at a rate of $\sim$ 300$M_\odot$ and old bursts ($\sim$ $10^8$ yr) at a lower star formation rate ($\sim$ 20 $M_\odot$) are consistent with the VLIRG nuclei 
distribution in the figure. However, old bursts (say, $> 10^7$ yr) require an extinction of 
$A_v \sim$ 1.5, somewhat smaller than young ones ($A_v \sim$ 2.5), the latter being in better 
agreement with the values inferred from spectroscopy. Under the hypothesis of an 
instantaneous burst, extinction should be rather small except for young bursts. For instance, 
a burst with less than 5$\times$ $10^6$ yr in age and $10^9M_\odot$ will require a mean extinction of 
$A_v \sim$ 2.5. Any other older burst will require smaller extinction and/or a larger 
mass.  Therefore, if the above mentioned mean reddening values inferred from the Balmer 
decrement are accepted, the present data favor young bursts ($< 5-10$ Myr) with a 
continuous rate of star formation of $\sim$ 300 $M_\odot$/yr, or an instantaneous burst of 
$\sim$ 10$^9M_\odot$.

\subsubsection {Comparison with the ULIRG samples: Are VLIRGs Low-redshift submillimeter galaxies ?}

In Figure 4,  we have also represented the values corresponding to the nuclei of the cool and warm ULIRGs studied by Surace et al (1998, 2000). Comparing the two types of ULIRG samples, it is clear that the warm ULIRGs show a much larger scatter in luminosity. In fact,
the warm ULIRGs extend $\sim$ 9 magnitudes in $M_B$, while the cool ULIRGs only $\sim$ 4, similar to the VLIRGs. The presence in the warm ULIRGs sample of the two low luminosity nuclei of IRAS 08572+3915 and, on the other end, the luminous Seyferts Mrk 1014 and Mrk 231 substantially increases the scatter. IRAS 08572+3915 has been recently studied in detail at optical wavelengths by means of integral field spectroscopy by Arribas and Colina (2000). If the nuclei positions in the color-magnitude diagram are corrected by the reddening inferred in that paper (i.e. $A_v \sim$ 2.4 and 0.6 for northern and southern nuclei, respectively) the scatter is not substantially reduced.  This suggests that the intrinsic scatter of the warm-ULIRG sequence is larger than both the cool-ULIRG and VLIRG samples.  

In any case, the VLIRG nuclei sequence agrees pretty well with both the cool and warm ULIRG nuclei sequences.  A good agreement is also shown between the VLIRG and the cool-ULIRG systems. The trend shown in Table 4 (in the sense that cool-ULIRGs have relatively stronger nuclei) is very weak if the scatter among individual objects is taken into account. Summarizing, the VLIRG and ULIRG samples, when  taken as a whole, are indistinguishable in the optical color-magnitude diagrams.

Recently, Frayer et al. (2003) have found that the observed near infrared colors of submillimeter-selected galaxies (SMG) are compatible with rest-frame optical colors and luminosities of low-redshifted ULIRGs, suggesting that nearby ULIRGs could be the low redshift analogues of SMGs. These results assume an average redshift of $\sim$ 2.5 for the submillimeter galaxy (SMG) population, according to the results of Chapman et al. (2003). The present data suggest that the rest-frame optical luminosities and colors of the SMGs are also consistent with those of the VLIRGs and, therefore, the optical connection between SMDs and ULIRGs can be extended to the more numerous population of VLIRGs. This raises the possibility that VLIRGs are also the low-redshift counterpart of SMGs. 

In Figure 4, we have also represented, for reference, the region where the star clusters around ULIRGs were detected (shaded area). Surace et al. (1998) found that these clusters typically have masses of $\sim$ $10^6 M_\odot$, values considerable lower than those inferred for the nuclei under the hypothesis that the latter are compact instantaneous starbursts (see above).   

\subsubsection {Dependence on Interaction Phase}

In Figure 5, we have broken down the samples of VLIRG and ULIRG systems and nuclei according to their interaction classification (see Section 4.1) in the $M_B - (B-I)$ diagram. As discussed before, we have only considered two interaction sub-classes: {\it young} systems, which are those with classes I, II, and III according to the Surace (1998) and Veilleux et al. (2002) classification scheme, and {\it old} systems, with classes IV and V. 

For the VLIRG systems, we observe a marginal trend in the sense that old systems tend to be redder and less luminous than young ones (also see Table 4). Just the opposite behavior is observed in the ULIRG system sequence. Older objects are bluer and, on average, more luminous. 
The nuclei show a similar trend. The mean magnitude and color values of VLIRG nuclei in young systems ($<M_B>$ = -17.90 $\pm$ 0.87, $<(B-I)>$= 2.15 $\pm$ 0.62) agree well with the corresponding values for the ULIRG nuclei ($<M_B>$ = -17.59 $\pm$ 1.47, $<(B-I)>$= 2.28 $\pm$ 0.60).  However, when old objects are compared, the nuclei of VLIRGs ($<M_B>$ = -17.72 $\pm$ 0.55, $<(B-I)>$= 2.49 $\pm$ 0.41) are less luminous and redder than the ones for the ULIRGs ($<M_B>$ = -18.80 $\pm$ 1.80, $<(B-I)>$= 1.70 $\pm$ 0.65).

To further analyze the behavior suggested by the mean values and also seen in Figure 5, we have used the task {\sc twosampt} in IRAF with the Gehan Generalized Wilcoxon two-sample test estimator to see if the different sub-classes are drawn from the same parent population.
If the young VLIRG and ULIRG systems are compared, this estimator gives a 63 percent chance that they are from the same population, so they are not substantially different. However, if the same test is run for old VLIRG and ULIRG systems, this estimator drops to 1 percent suggesting that, according to their photometric values, they belong to two distinct populations. For the nuclei we found a similar trend. The comparison of nuclei in young ULIRGs and VLIRGs gives a 72 percent chance that they are from the same parent population. However, the comparison between the nuclei of the old ULIRGs and VLIRGs gives only a 2 percent likelihood. Although we cannot rule out that these test results are affected by selection effects and the relatively low numbers in the samples, the magnitude of the difference between young and old classes (63 versus 1, and 72 versus 2) suggests that there is a clear difference between the optical photometric properties of old VLIRGs and ULIRGs. 

In any case, if confirmed with larger samples, the observed behavior in the color-magnitude diagram indicates that the evolution during the late phases is different for VLIRGs and ULIRGs. This is consistent with other works which have suggested a correlation of type of nuclear ionization (i.e AGN/HII) with bolometric (far infrared) luminosity (e.g. Veilleux et al. 2002, and references there in). Since the evolution of an instantaneous  compact luminous starburst in the color-magnitude diagram goes in the direction of producing redder colors and lower absolute magnitudes (as opposed to a continuous burst for which the luminosity increases with time; see Figure 3), the observed behavior is consistent with the idea that VLIRG activity tends to be dominated by instanteneous starbursts, while a higher proportion of ULIRGs may evolve into a QSO-like object.  

\section{Conclusions}

We have obtained multi-wavelength optical imaging and photometry of a sample of 19 low redshift (0.03$<$z$<$0.07) very luminous infrared galaxy (VLIRG) systems. These objects have morphological characteristics similar to those found in ULIRGs suggesting strong interactions and/or mergers. The main conclusions of the present study are:

1)  In the color-magnitude diagram, the nuclei and the systems have well defined distributions which run along the reddening line, suggesting that part of the observed scatter is due to extinction effects. Most of the nuclei have redder colors than the galaxy disks, though there are a few exceptions. Typically, the nuclei comprise 10 percent of the total flux of the system in B, and 13 percent in I. Considering the mean extinction values from the Balmer decrement given by Veilleux et al., the mean intrinsic magnitudes of the nuclei in the VLIRG sample are: $< M_B>_o$ = -22.75 and $<M_I>_o$ = -21.90 (i.e. $(B-V)_o$ = - 0.85).

2) The optical colors and magnitudes for the nuclei and the systems of the VLIRGs agree well with those for the cool ULIRGs. The VLIRG nuclei also agree with those of the warm ULIRGs, though the latter show a much larger scatter in both luminosity and color.  Despite the difference in bolometric/infrared luminosity, the photometric properties of the VLIRG and ULIRG samples, considered as a whole, are indistinguishable at optical wavelengths (i.e. morphologies, compactness, magnitudes, and colors). 

3) The recent suggestion by Frayer et al., based on near infrared photometry, that nearby ULIRGs could be the local analogue of submillimeter galaxies (SMGs) can also be extended to the more numerous population of VLIRGs. In fact, the previous conclusion suggests that the optical luminosities and colors of VLIRGs are also in good agreement with the corresponding rest-frame values of the SMGs, reinforcing the connection low-{\it z} LIRGs -- high-{\it z} SMGs. 

4) When considering {\it young} and {\it old} systems separately, according to the interaction classification scheme proposed by Surace (see also Veilleux et al. 2002), some differences are found between the VLIRG and ULIRG samples. In particular, although the young VLIRGs and ULIRGs seem to share similar properties, the old VLIRGs are less luminous and redder than old ULIRG systems. If confirmed with larger samples, this behavior suggests that the late evolution of VLIRGs and ULIRGs is different. 

5) Under the starburst scenario, the activity in the nuclei of the VLIRGs could be, in principle, explained by young bursts of less than $10^7$ years in age with a continuous rate of star formation of $\sim$ 300 $M_\odot$/yr), or with an instantaneous creation of $\sim$ 10$^9M_\odot$. However, the observed behavior, according to which old systems have lower luminosities and redder colors, is in better agreement with the evolution of an instantaneous starburst. Our data are also consistent with the idea that VLIRG activity tends to be dominated by starbursts, while a higher proportion of ULIRGs may evolve into a QSO-like object.

\acknowledgements

This paper is based on observations made with the Nordic Optical Telescope, operated
on the island of La Palma jointly by Denmark, Finland, Iceland, Norway,
and Sweden, in the Spanish Observatorio del Roque de los Muchachos of
the Instituto de Astrofisica de Canarias. The data presented here have
been taken using ALFOSC, which is owned by the Instituto de Astrofisica
de Andalucia (IAA) and operated at the Nordic Optical Telescope under
agreement between IAA and the NBIfAFG of the Astronomical Observatory
of Copenhagen.  We thank Dave Clements, Jes\'us Ma\'\i z-Apell\'aniz, and Ed Nelan 
for providing us with useful information. This work was partially funded by the 
STScI - Director's Discretionary Research Fund (DDRF) program.

\appendix

\section{Appendix: Individual Objects}

{\bf\underline {IRAS 14071-0806:}} 
The central object looks like two edge-on disk galaxies crossing each
other.  Two nuclei separated  by 4.3 ''(4.5 kpc) are clearly detected
in $B, V,$ and $ I$.  The eastern nucleus is brighter in the three filters.
These nuclei are in an asymmetric location with respect to the
low-intensity envelope (putative disks) of the two galaxies. A
relatively red knot is found southwest of the brightest nucleus, at 7.1
'' (7.2 kpc) from it. A 'bridge' of low emission (tidal tail) seems to connect
(in projection)  the southern part of these galaxies with a  bright  object
located  24'' (25 kpc)  away along PA 150 degrees.
However,  this is likely  coincidental since this object has a star-like intensity profile.  A faint tidal tail extending from the northern nucleus
towards the SE (PA about 120 degrees) with a projected size of 76 kpc
is clearly detected in $B$ and  $V$. However it is not observed in our $I$ image.

{\bf\underline {IRAS 16104+5235 (NGC6090):}} 
Merging of two face-on galaxies.  The nuclei are at a projected
distance of 5.5'' (3.1 kpc) and are clearly detected. The morphology in the
innermost regions around the nuclei (1-2 kpc) is irregular. A 'bridge'
connects the northern part of these two regions. A common envelope
extends up to about 11.3'' (6.4 kpc).  Within this envelope several
knots (especially towards the NE) are detected.  Two huge tidal tails extend at least 65 kpc to the south
and 50 kpc NE. The field is full of what look like clusters or dwarf galaxies,
some of which could have been involved in the merger. This object is included in
the sample by Scoville et al. (2000).

{\bf\underline {IRAS 16180+3753 (NGC 6120):}} 
Evidence for a faint tidal tail extending towards the north up to at
least 37 arcsec (22 kpc) from the center. This tidal tail seems more
prominent at longer wavelengths (i.e. $I$-band image). Some knots seem to be associated
with this tidal tail. In a symmetrical position with respect to this
tidal tail, two faint patches could indicate the remains of a SE tidal
tail. In $B$, the innermost regions have a rather irregular structure with
three local maxima, which suggests that the flux is dominated by
chains of clusters similar to those found in some ULIRGs. However, the peak 
in $I$ is rather prominent and well centered with the outer envelope, 
with a secondary peak towards the W. This secondary peak 
is the brightest region in $B$.  The common egg-like envelope extends 
more than 30 kpc in its largest dimension.

{\bf\underline {IRAS 16396+7814:}}
The system seems to be formed by two main galaxies, which are more than
50 kpc apart. These objects are clearly connected by a  tidal
tail. In projection, a third galaxy is in the vicinity (69 kpc NE from
the brightest galaxy) though there is no evidence for a physical
connection with either of the other two. However, a faint structure at 34
kpc NE could be a dwarf galaxy or a cluster associated with the system.
The main object looks like a face-on disrupted spiral. The flux peak is well centered with the inner structures. A secondary peak
is found at 6.4'' (6.75 kpc) north from the main peak.

{\bf\underline {IRAS 16569+8105:}}
No evidence for more than one object or nucleus. Irregular inner light
distribution within an egg-like low intensity envelope of 25.6 '' (24
kpc). The $I$ image reveals a spiral structure similar to that of the
ULIRG IRAS 15206+3342 (Arribas and Colina, 2002). However, there is no
evidence for tidal tails.

{\bf\underline {IRAS 16577+5900 (NGC6286 + NGC6285):}}
The system consists of two disk galaxies separated by a projected
distance of 32 kpc. A faint bridge of emission between the galaxies is a
clear sign of this interaction. The galaxy located at the SE (NGC6286)
looks like an edge-on disk galaxy. The inner parts of the disk are
clearly disrupted by the interaction. Towards the SE, a separated region
of relatively strong emission seems to be part of the tidal tail
connecting the two galaxies. The galaxy at the NW (NGC 6285) has a
spiral structure. Its inner 7 kpc shows a rather inhomogenous and
irregular structure, likely due to the effects of the internal
extinction, as well as the presence of several knots (clusters),
especially towards the West. The spiral-arm at the NW seems to connect
with a tidal tail whose onset is at the south of the nucleus and bends
in the same direction as the spiral arm.

{\bf\underline {IRAS 17366+8646 (Mrk 1116):}}
Two galaxies in interaction separated by 42'' (44 kpc). The largest
is a face-on spiral rotating clockwise. The northern arm connects with
the smaller galaxy. The other spiral arm extends towards the south
where a patchy structure (dwarf galaxy ?) is clearly visible in the
three filters. Another galaxy is found at 60$''$ (62 kpc) SE, but it is
unclear if this object is connected with IRAS 17366+8646.

{\bf\underline {IRAS 17487+5637:}}
The low intensity isophotes of this galaxy are elliptical in shape
(major axis PA $\sim$ 135 degrees), while the brightest/inner regions are
more irregular. The center of these inner isophotes (nucleus ?) is
off-set with respect to the outer isophotes by about 2.5'' (3.4 kpc).
The inner structure also is similar to the spiral-like structure of the ULIRG
IRAS 15206+3342 (Arribas and Colina, 2002). At 43'' (55 kpc) north from 
IRAS 17487+5637 there is another galaxy more irregular in shape. Our
images do not show a clear sign of interaction, though this possibility
cannot be ruled out.

{\bf\underline {IRAS 18329+5950 (NGC6670):}}
The low intensity isophotes have a banana-like morphology. The inner
and  brighter isophotes reveal the presence of two nearly edge-on disk
galaxies of similar optical luminosity. The system extends at least
70'' (40 kpc). There is no evidence for a tidal tail, but there is low surface brightness extended emission all around it.

{\bf\underline {IRAS 18432+6417:}}
The system looks like an inclined disk (of high ellipticity). The
nucleus in the $I$ image is well centered with the outer isophotes. In
the $B$ image, the position of the nucleus has a local minimum, which
suggests the presence of high internal extinction there.

{\bf\underline {IRAS 19171+4707:}}
The low intensity isophotes have an elliptical shaped morphology,
though the presence of a spiral arm is rather clear. Again, this
morphology is reminiscent  of the ULIRG IRAS 15206+3342 (Arribas and Colina,
2002). The nucleus is off-set with respect to the outer isolines by 1.9
arcsec (2 kpc). No evidence for tidal tails is found. Towards the SE, there
is an elliptical at 50'' (54kpc) and a spiral at 100'' (110 kpc), but there
is no evidence for interaction with IRAS 119171+4707.

{\bf\underline {IRAS 19354+4559:}}
The whole system has an irregular morphology. It likely consists of two
merging disk galaxies whose nuclei are separated by about 8.5 arcsec
(10.6 kpc). One looks like an edge-on disk, while the morphology of the
other is more uncertain due to the presence of several likely field
stars. However, the $I$ image suggests this is a face-on disk rotating
clockwise.

{\bf\underline {IRAS 19545+1625:}}
The presence of many field stars complicates the morphological study of
this galaxy. It looks like an isolated single object with a elliptical shape
at low intensities, and a more complex structure inwards with perhaps
the presence of a spiral arm.

{\bf\underline {IRAS 20135-0857:}}
The system consists of two galaxies close in projection at 16.4''
(12 kpc). However, our images do not reveal clear signs of interaction
among them (for instance, bridges or tidal tails). The galaxy at the NE is
relatively more prominent in $I$, which likely indicates it has more
mass. The low intensity isophotes suggest the presence of somewhat distorted spiral arms, with
a more complicated structure inwards. The secondary galaxy could also be
a disrupted elliptical, though the presence of field stars may
also hide the galaxy's actual morphology.

{\bf\underline {IRAS 20210+1121:}}
Two galaxies separated by 12.2'' (13.3 kpc) in clear interaction. The
southern galaxy shows a prominent spiral arm structure, while the
northern object is more spheroidal in shape. No large tidal tails
beyond the common envelope are detected, though a bridge of emission
connecting both galaxies is rather conspicuous.

{\bf\underline {IRAS 20550+1656:}}
Irregular morphology, with several emitting regions. The $I$ image reveals
the presence of two relatively bright regions (separated by 11.5'', 8
kpc) within the most intense structure, which could harbor the nuclei
of the two putative merging galaxies.  A fainter region  is located at
22'' (15.5 kpc) from the brightest nucleus. The overall low-intensity
envelope has a banana-like morphology.

{\bf\underline {IRAS 21048+3351:}}
Two nearly edge-on disk galaxies of similar optical luminosity crossing
each other. Their nuclei are at a projected distance of 8.9 kpc
(9'').  A long tidal tail extends  at least 36 '' ( 35 kpc) in the SW
direction.

{\bf\underline {IRAS 21248+2342:}}
The system is elliptical in shape (major axis PA about 135 degrees), which could be the result of an inclined disk.
The nucleus in the $I$ image is well centered with the outer isophotes.
A local maximum is found towards the NW, which is coincident with the
nucleus in the $B$ image. This is likely due to the effects of internal
reddening, though it could also be produced by contamination from a field
star. No tidal tails are detected.

{\bf\underline {IRAS 22357-1702:}}
The system has a rather regular elliptical morphology, though the $B$
image suggests the presence of a spiral-arm structure inwards. There is
no evidence for the presence of two galaxies (or nuclei).

\clearpage

\begin{deluxetable}{crrrrrrrrrrrr}
\tabletypesize{\scriptsize}
\tablecaption{Properties of the Sample \tablenotemark{a}\label{tbl-1}}
\tablewidth{0pt}
\tablehead{
\colhead{IRAS-ID} & 
\colhead{$\alpha$(J2000) }   & 
\colhead{$\delta$(J2000) }   &
\colhead{$z$} &
\colhead{$f_{12}$}  & 
\colhead{$f_{25}$} & 
\colhead{$f_{60}$} &
\colhead{$f_{100}$} & 
\colhead{$D_L$}  &
\colhead{$L_{60}$}   & 
\colhead{$L_{ir}$}           &
\colhead{$f_{25}/f_{60}$} &
\colhead{$Ref.$}\\

    & hh:mm:ss     & $\deg$:$'$:$''$  &   & Jy & Jy & Jy & Jy  & Mpc &  &  &   &    \\
}
\startdata

 14071-0806 & 14:09:46.60 & -08:20:06.0 & 0.05345 & $<$0.16 & $<$0.35 & 1.59  & 2.06  & 205.45 & 11.12 &$<$ 11.38 &$<$0.22 & F95 \\
 16104+5235\tablenotemark{b} & 16:11:40.70 & +52:27:24.0 & 0.02930 & 0.26    & 1.11    & 6.66  & 8.94  & 114.61 & 11.22 & 11.41 & 0.17 & S92 \\
 16180+3753\tablenotemark{b} & 16:19:48.10 & +37:46:27.9 & 0.03059 & 0.24    & 0.50    & 3.99  & 8.03  & 119.54 & 11.03 & 11.28 & 0.13 & S92 \\
 16396+7814 & 16:37:20.50 & +78:08:28.0 & 0.05450 & 0.06    & 0.15    & 1.20  & 2.53  & 209.33 & 11.02 & 11.25 & 0.13 & F95 \\
 16569+8105 & 16:52:34.50 & +81:00:18.0 & 0.04915 & 0.09    & 0.17    & 1.56  & 3.73  & 189.50 & 11.04 & 11.29 & 0.11 & S92 \\
 16577+5900\tablenotemark{b} & 16:58:31.65 & +58:56:14.3 & 0.01835 & 0.33    & 0.49    & 7.88  & 22.6  & 72.35  & 10.88 & 11.16 & 0.06 & S92 \\
 17366+8646\tablenotemark{b} & 17:19:28.42 & +86:44:19.0 & 0.02636 & 0.31    & 0.49    & 4.70  & 9.73  & 103.33 & 10.97 & 11.23 & 0.10 & S92 \\
 17487+5637 & 17:49:37.40 & +56:37:06.7 & 0.06567 & 0.07    & 0.18    & 1.60  & 4.01  & 250.24 & 11.31 & 11.54 & 0.11 & S92 \\
 18329+5950\tablenotemark{b} & 18:33:35.15 & +59:53;21.0 & 0.02885 & 0.36    & 1.03    & 8.25  & 15.2  & 112.88 & 11.30 & 11.52 & 0.12 & S92 \\
 18432+6417 & 18:43:30.20 & +64:20:57.0 & 0.07409 & $<$0.10 & 0.04    & 0.96  & 2.05  & 280.65 & 11.19 &$<$ 11.43 & 0.04 & S95 \\
 19171+4707 & 19:18:33.29 & +47:13;14.0 & 0.05563 & $<$0.25 & $<$0.25 & 0.87  & 2.74  & 213.50 & 10.90 &$<$ 11.39 & $<$0.29 & F95 \\
 19354+4559 & 19:36:59.90 & +46:06:28.0 & 0.06480 & $<$0.77 & $<$0.25 & 0.69  & 1.69  & 247.07 & 10.93 &$<$ 11.71 & $<$ 0.36 & L99 \\
 19545+1625 & 19:56:51.12 & +16:33:39.0 & 0.03931 & $<$0.25 & $<$0.25 & 2.03  & 4.98  & 152.64 & 10.96 &$<$ 11.29 & $<$ 0.12 & S92 \\
 20135-0857\tablenotemark{c} & 20:16:17.73 & -08:47:43.8 & 0.05741 & 1.15    & 0.52    & 1.26  & 2.44  & 220.05 & 11.09 & 11.81 & 0.41 & F95 \\
 20210+1121\tablenotemark{c} & 20:23:25.40 & +11:31:34.0 & 0.05639 & 0.29    & 1.40    & 3.39  & 2.68  & 216.30 & 11.50 & 11.77 & 0.41 & S92 \\
 20550+1656\tablenotemark{b} & 20:57:23.29 & +17:07:34.3 & 0.03610 & 0.25    & 2.39    & 13.3  & 10.6  & 140.51 & 11.70 & 11.82 & 0.18 & S92 \\
 21048+3351 & 21:06:53.26 & +34:04:00.8 & 0.04962 & $<$0.28 & $<$0.25 & 1.82  & 2.38  & 191.25 & 11.12 &$<$ 11.39 & $<$0.14 & S92 \\
 21248+2342 & 21:27:03.30 & +23:55:45.0 & 0.05096 & $<$0.25 & $<$0.25 & 0.94  & 2.70  & 196.23 & 10.85 &$<$ 11.32 & $<$0.27 & S95\\
 22357-1702 & 22:38:25.49 & -16:46:48.2 & 0.05650 & $<$0.12 & $<$0.27 & 1.13  & 1.99  & 216.70 & 11.02 &$<$ 11.32 & $<$ 0.24 & S95 \\

 \enddata


\tablenotetext{a}{NOTES- Column 1: IRAS identification. Columns 2-3: Coordinates from the NED. Column 4: redshift from source indicated in column 13. Columns 5-8: flux densities from the IRAS Faint Source Catalog. Column 9: $D_L$, Luminosity distance for the Einstein-de Sitter universe with $H_o$=75 km/s/Mpc. Column 10: $L_{60}$, logarithm of the luminosity in units of $L_\odot$ computed using the 60 $\mu$m flux densities from the IRAS Faint Source Catalog. Column 11: $L_{ir}$, logarithm of the luminosity in units of $L_\odot$ computed using the prescription given in Sanders and Mirabel (1996). Column 12: Far infrared color. Column 13: redshift source: S92, Straus et al. 1992; F95, Fisher et al. 1995; S95, Straus et al. 1995 (obtained from the NED); L99, Lawrence et al. 1999.}
\tablenotetext{b}{Optical spectroscopic data available in Kim et al. (1995), Veilleux et al. (1995), or Wu et al. (1998b)}
\tablenotetext{c}{'Warm' VLIRG}
\end{deluxetable}

\clearpage

\begin{deluxetable}{crrrrr}
\tabletypesize{\scriptsize}
\tablecaption{Observations \label{tbl-2}}
\tablewidth{0pt}
\tablehead{
\colhead{IRAS-ID} & 
\colhead{Scale} & 
\colhead{Filters (Exp. t)}   &
\colhead{Air mass} &
\colhead{Seeing \tablenotemark{a} } & 
\colhead{Date}\\
            & pc/$''$ &     (sec)                &         &   arcsec   &  \\
}
\startdata

14071-0806  & 996 & B(500), V(500), I(400) & 1.38, 1.57, 1.89&   0.94	&  8-July-02  \\
16104+5235  & 555 & B(400), V(300), I(300) & 1.11, 1.09, 1.10&   0.94	& 10-July-02 \\
16180+3753  & 579 & B(400), V(300), I(300) & 1.02, 1.02, 1.01&   1.20	& 11-July-02 \\
16396+7814  & 1015 & B(400), V(300), I(300) &1.53, 1.54, 1.56&   1.28	& 11-July-02 \\
16569+8105 & 918 & B(400), I(300) & 1.67, 1.70 &  1.20\tablenotemark{b} 	& 11-July-02 \\
16577+5900  & 350 & B(400), V(300), I(300) &1.17, 1.17, 1.16 &   0.75	&  9-July-02 \\
17366+8646  & 501 & B(300), V(300), I(300) & 2.00, 1.99, 1.95&  0.98\tablenotemark{b} 	&  8-July-02 \\
17487+5637  & 1213 & B(400), I(300) & 1.15, 1.13 &  0.75  & 11-July-02 \\
18329+5950  & 548 & B(300), V(300), I(300) & 1.17, 1.19, 1.21&  0.75 &  8-July-02 \\
18432+6417  & 1360 & B(400), I(300) & 1.24, 1.23 &  0.67 &  9-July-02 \\
19171+4707  & 1035 & B(400), I(300) & 1.06, 1.05&  0.67 &  9-July-02 \\
19354+4559  & 1199 & B(400), I(300) & 1.05, 1.05&   0.60 & 10-July-02 \\
19545+1625  & 740 & B(400), I(300) & 1.03, 1.02 &   0.94 & 11-July-02 \\
20135-0857  & 1067 & B(400), I(300) & 1.26, 1.28&   0.90   &  9-July-02 \\
20210+1121  & 1049 & B(400), V(300), I(300) & 1.05, 1.07, 1.04	&   0.75	& 10-July-02 \\
20550+1656  & 679 & B(300), V(300), I(300) & 1.06, 1.11, 1.17&  0.98	&  8-July-02 \\
21048+3351  & 928 & B(400), I(300) & 1.00, 1.00 &   0.82 & 11-July-02 \\ 
21248+2342  & 951 & B(400), I(300) & 1.00, 1.00 &   0.75	&  9-July-02 \\
            & & B(400), V(300), I(300) & 1.02, 1.03, 1.05&  0.75	& 10-July-02 \\
22357-1702  & 1051 & B(400), I(300) & 1.42, 1.44 &   1.3 & 11-July-02 \\
m92         & & B(180), V(120), I(120) & 1.074, 1.058, 1.041  &      &  8-July-02 \\
m92         & & B(180), V(120), I(120) & 1.035, 1.032, 1.031  &      &  9-July-02 \\
m92         & & B(180), V(120), I(120) & 1.037, 1.042, 1.048 &      & 10-July-02 \\

\enddata
\tablenotetext{a}{FWHM of stars' images, measured in the $I$ images}
\tablenotetext{b}{Egg-shaped images, probably due to guiding instabilities}
\end{deluxetable}



\clearpage

\begin{deluxetable}{lrrrrcrrrrrr}
\tabletypesize{\scriptsize}
\tablewidth{0pt}
\tablecaption{Photometry}
\tablehead{
\colhead{IRAS (NGC) Object /Region\tablenotemark{k}}           & \colhead{Class}      &
\colhead{ $\Delta\alpha ('') $}          & \colhead{$\Delta\delta ('') $}  &
\colhead{$M_B$}          & \colhead{$B$}    &
\colhead{$V$}  & \colhead{$I$}  &
\colhead{$B-V$} & \colhead{$B-I$} &
\colhead{$V-I$} &

}
\startdata

14071-0806            & III   & & & & &                     &         &      &  & \\
\hline

System   &  & & & -20.29& 16.38 &  15.72 &  14.73 &   0.67 &   1.65 &   0.98 & \\
Galaxy 1/Total\tablenotemark{a}& & 0.0& 0.0& -19.79& 16.88 &  16.17 & 15.16  &  0.71  &  1.71 &   1.00 \\
Galaxy 1/Nucleus& & & & -18.60& 18.07 &  17.52 &  16.56 &   0.55 &   1.52  &  0.96 & \\
Galaxy 2/Total\tablenotemark{a} & & -3.8 & 0.6& -19.22& 17.45 &  16.73 &  15.93 &   0.72 &   1.51  &  0.79 \\
Galaxy 2/Nucleus& & & &-17.84& 18.83 &  18.18 &  17.12 &   0.65 &   1.71  &  1.06 \\
\hline
16104+5235 (NGC 6090)     & III      & & & & &                    &         &      &  &\\
\hline
System & &  & &-20.97& 14.36 &  13.83 &  12.99 &   0.53 &   1.37  &  0.84\\
Galaxy 1/Total & & 0.0 & 0.0 &-20.44& 14.90 &  14.28 &  13.39 &   0.61 &   1.51  &  0.89 \\
Galaxy 1/Nucleus & & & &-19.49& 15.87 &  15.31 &  14.44 &   0.56 &   1.43  &  0.86 \\
Galaxy 2/Total & & -5.1 & -4.9 & -19.94& 15.40 &  15.01 &  14.26 &   0.38 &   1.14 &   0.76 \\
Galaxy 2/Nucleus & & & & -19.23& 16.13 &  15.83 &  15.20 &   0.30 &   0.93  &  0.63 \\
\hline
16180+3753 (NGC 6120)     & V       & & & & &                    &         &      &  & \\
\hline
System/galaxy & & & & -21.14& 14.35 &  13.64 &  12.59 &   0.71 &   1.76  &  1.05 \\
Nucleus \tablenotemark{b} & &0.0 &0.0 &-18.29 & 17.16  &  16.23 &  14.86  &  0.94  &   2.31 &  1.37  \\
Secondary Nucleus\tablenotemark{c} & &-4.6 &-0.1 &-18.62& 16.83 &  16.41 &  15.59 &   0.42 &   1.24  &  0.82 \\
\hline
16396+7814    & III      & & & & &                    &         &      &  & \\
\hline
 System\tablenotemark{d} &  & & &-21.93& 14.78 &  13.99 &  12.96 &   0.79 &   1.83  &  1.03 \\
 Galaxy 1/Total & &0.0 &0.0 &-21.76& 14.95 &  14.17 &  13.15  &  0.78 &   1.81  &  1.03 \\
 Galaxy 1/Nucleus & &  & &-17.52& 19.20 &  18.07 &  16.63 &   1.13  &  2.58  &  1.44 \\
 Galaxy 2/Total & &-3.0 &51.7 & -19.80& 16.90 &  16.05 &  14.96 &   0.85  &  1.93  &  1.09 \\
 Galaxy 2/Nucleus & & & & -17.73& 18.99 &  18.17 &  16.83 &   0.83  &  2.16  &  1.34 & \\
 Galaxy 3/Total\tablenotemark{e} & &31.0 &56.8 & -20.1& 16.61 &  15.48 &  14.24 &   1.13 &   2.38  &  1.25 \\
 Galaxy 3/Nucleus & & & & -17.94& 18.78 &  17.56 &  16.23 &   1.22 &   2.55  &  1.33 \\
\hline
 16569+8105   & V       & & & & &                    &         &      &  & \\
 \hline
System/galaxy&  & & & -20.73& 15.76 & \nodata  &  14.09 & \nodata    & 1.67   & \nodata\\
Nucleus & & & & -18.26&  18.23 & \nodata  &  16.26 & \nodata   &  1.97  & \nodata\\
\hline
16577+5900 (NGC6285/6)    & III      & & & & &                   &         &      &  & \\
\hline
 System & & & &-20.16& 14.14 &  13.29 &  12.05 &   0.86  &  2.10  &  1.24 \\
 Galaxy 1/Total & &0.0  &0.0 &-19.75& 14.55 &  13.70 &  12.49 &   0.85  &  2.06  &  1.21\\
 Galaxy 1/Nucleus & & & &-17.64& 16.69 &  15.79  & 14.42 &   0.90  &  2.27  &  1.37\\
 Galaxy 2/Total & &-57.9 &69.6 & -19.28& 15.02 &  14.36 &  13.23 &   0.66  &  1.79  &  1.13\\
 Galaxy 2/Nucleus & & & & -17.77& 16.56 &  15.82 &  14.65 &   0.74  &  1.91  &  1.17 \\
\hline
17366+8646    & III        & & & & &                    &         &      &  & \\
\hline
System & &&  & -21.16& 13.93 &  13.32 &  12.32 &   0.61 &   1.61  &  1.00 \\
Galaxy 1/Total& & 0.0&0.0 & -21.01& 14.09 &  13.53 &  12.59 &   0.56  &  1.50   & 0.94\\
Galaxy 1/Nucleus& & & & -18.59&16.53 &  15.81 &  14.65 &   0.72  &  1.88  &  1.16\\
Galaxy 2/Total& & 41.3&-6.8 & -18.95&16.15 &  15.23 &  13.95 &   0.92  &  2.20  &  1.28\\
Galaxy 2/Nucleus& & & &-17.52& 17.60 &  16.60  & 15.21 &   1.00  &  2.40  &  1.40\\
\hline
17487+5637   & V\tablenotemark{g}        & & & & &                    &         &      &  & \\
\hline
Galaxy 1/ System & &0.0 &0.0 & -20.33&16.82 & \nodata &  14.94 & \nodata &  1.88  & \nodata \\
Galaxy 1/Nucleus& & & & -18.12& 19.01 & \nodata &  17.00  &\nodata  &  2.01  & \nodata \\
Galaxy 2/ Total \tablenotemark{e} && 3.0 &41.7 &-19.63& 17.52 & \nodata  &  15.96 & \nodata &  1.56  & \nodata\\
Galaxy 2/Nucleus& & & & -16.70& 20.43 &\nodata  &  18.50 &  \nodata &  1.93  & \nodata\\
\hline
18329+5950 (NGC 6670)      & III     & & & & &                   &         &      &  & \\
\hline
System && & & -20.62& 14.72 &  13.93 &  12.75 &   0.79 &   1.97  &  1.18\\
Galaxy 1/Total & &0.0 &0.0 & -20.03& 15.31 &  14.56 &  13.53 &   0.75  &  1.77  &  1.02\\
Galaxy 1/Nucleus& & & & -17.90& 17.42 &  16.55 &  15.18 &   0.87  &  2.24  &  1.37\\
Galaxy 2/Total & & -27.5 &-5.7 & -19.52& 15.82 &  14.92 &  13.74 &   0.91  &  2.09  &  1.18\\
Galaxy 2/Nucleus& & & & -16.94& 18.38 &  17.26 &  15.53 &   1.13  &  2.85  &  1.73\\
\hline
18432+6417    & V\tablenotemark{h}        & & & & &                    &         &      &  &\\
\hline
Galaxy/System & & & & -19.93& 17.48 & \nodata  &  15.24  & \nodata  &   2.23  & \nodata\\
Nucleus/Nucleus& & & & -17.07& 20.32 & \nodata  &  17.56  &  \nodata  &  2.75  &\nodata \\
\hline
19171+4707     & V      & & & & &                    &         &      &  & \\
\hline
Galaxy/System& & & & -20.61& 16.18 & \nodata  &  14.25 & \nodata  &  1.93  & \nodata\\
Nucleus/Nucleus& & & & -18.02& 18.74 & \nodata  &  16.34 & \nodata  &  2.40  &\nodata \\
\hline
19354+4559    & III       & & & & &                    &         &      &  &\\
\hline
 System & & & & -20.21& 16.91 & \nodata &  14.57 & \nodata  &  2.34  &\nodata \\
 Galaxy 1/Total & &0.0 &0.0 &-19.42& 17.70 & \nodata &  15.24 & \nodata   &  2.46  & \nodata\\
 Galaxy 1/Nucleus& & & & -17.28& 19.82 & \nodata  &  16.90  & \nodata  &  2.92  & \nodata\\
 Galaxy 2/Total & & 6.7 &5.3 & -19.52& 17.60 & \nodata &  15.33 &  \nodata  &  2.27  & \nodata\\
 Galaxy 2/Nucleus& & & & -16.78& 20.32 & \nodata &  17.35 & \nodata   &  2.97  & \nodata\\
\hline
19545+1625   & V         & & & & &                    &         &      &  & \\
\hline
Galaxy/System & & & & -19.15& 16.84 & \nodata &  14.11  & \nodata &  2.73  & \nodata\\
Galaxy/Nucleus&  & & &-16.98& 19.02 & \nodata &  15.94 & \nodata &  3.08  &\nodata\\
\hline
20135-0857    & I\tablenotemark{i}       & & & & &                    &         &      &  &\\
\hline
System & & & & -20.30 & 16.53 & \nodata & 14.48 & \nodata & 2.05 & \nodata \\
Galaxy 1/Total & &0.0  & 0.0&-19.86& 16.97 &\nodata   &  14.75 &\nodata   &  2.22  & \nodata\\
Galaxy 1 /Nucleus&  & &  &-17.61& 19.22 & \nodata  &  16.67 & \nodata  &  2.55  & \nodata\\
Galaxy 2/Total\tablenotemark{f}&&-14.5 & - 7.1 & -18.85& 17.98 & \nodata  &  16.57 & \nodata  &  1.41  &\nodata \\
Galaxy 2/Nucleus&  & & &-16.99& 19.84 & \nodata &  18.19 &  \nodata & 1.65   & \nodata\\
\hline
20210+1121   & III        & & & & &                    &         &      &  &\\
\hline
System &  & & &-20.94& 15.85 &  14.82 &  13.82  &  1.03  &  2.03  &  1.00 \\
Galaxy 1/Total & &0.0  &0.0 &-20.65& 16.14 &  15.11 &  14.20  &  1.03  &  1.94  &  0.91\\
Galaxy 1/Nucleus& &  & &-19.23& 17.56 &  16.33 &  15.79  &  1.22  &  1.77  &  0.54 \\
Galaxy 2/Total & & -3.0 &11.8 &-19.59& 17.20 &  16.15 &  14.80  &  1.04  &  2.40  &  1.36 \\
Galaxy2/Nucleus& &  & &-18.19& 18.60 &  17.48  & 16.19  &  1.12 &   2.41  &  1.29 \\
\hline
20550+1656 \tablenotemark{j}  & III        & & & & &                    &         &      &  & \\
\hline
Galaxy/System& &  & &-21.37& 14.44 &  13.93 &  13.14  &  0.52  &  1.30 &   0.79 \\
Nucleus&  & & &-19.62& 16.19 &  15.79 &  15.30 &   0.40  &  0.89 &   0.49 \\
\hline
21048+3351   & II     &  & & & &                    &         &      &  & \\
\hline   
System & & & &-19.39& 17.15 & \nodata &  14.58 & \nodata & 2.56  &\nodata \\
Galaxy 1/Total& & 0.0&0.0 & -18.45& 18.09 & \nodata &  15.35  & \nodata &  2.73  & \nodata\\
Galaxy 1/Nucleus& & & & -17.09& 19.42 & \nodata &  16.53  & \nodata &  2.88  & \nodata\\
Galaxy 2/Total& &-8.6 & 1.9 &-18.74& 17.80 & \nodata &  15.06 & \nodata &  2.73  & \nodata\\
Galaxy 2/Nucleus& &  & & -16.46& 20.05 & \nodata&  17.06  &\nodata &  2.98  & \nodata\\
\hline
21248+2342    & V       & & & & &                   &         &      &  & \\
\hline
System/Galaxy& &  & &-19.72& 16.86 &  15.94 &  14.60 &   0.92 &  2.26   & 1.34\\
Nucleus&  & & &-17.05& 19.52 &  18.25 &  16.44 &   1.27  &  3.08  &  1.81 \\
\hline
22357-1702     & V      & & & & &                    &         &      &  & \\
\hline
System/Galaxy &&  & & -20.48&16.31 & \nodata &  14.41  & \nodata  &  1.91  & \nodata \\
Nucleus& &  & & -18.03 & 18.77 & \nodata &  16.45 &  \nodata &  2.33  & \nodata \\
\enddata
\tablenotetext{a}{Uncertain extraction}
\tablenotetext{b}{Brightest region in $I$, well centered}
\tablenotetext{c}{Brightest region in $B$}
\tablenotetext{d}{Includes galaxies 1,2, and 4}
\tablenotetext{e}{Likely not part of the system}
\tablenotetext{f}{No evidence of interaction}
\tablenotetext{g}{This classification assumes that galaxy 2 is not part of the system}
\tablenotetext{h}{Only one galaxy obviously visible, but it is not clear that it has evidence for disturbance}
\tablenotetext{i}{Two disturbed galaxies in the field of view without apparent contact}
\tablenotetext{j}{The location of the secondary nucleus is uncertain, but it could be at -6.8$''$, +9.6$''$}
\tablenotetext{k}{Nuclear photometry refers to a 2.5 kpc aperture}

\end{deluxetable}

\clearpage

\begin{deluxetable}{crrrrrrr}
\tabletypesize{\scriptsize}
\tablecaption{Mean Photometric Properties\tablenotemark{a} \label{tbl-4}}
\tablewidth{0pt}
\tablehead{
\colhead{Group}& 
\colhead{n} &
\colhead{$<z>$} &
\colhead{$<M_B>$} & 
\colhead{$<M_I>$}   &
\colhead{$<{\it B-I}>$} &
\colhead{$< L_n/L_s |_{\it B}>$} & 
\colhead{$< L_n/L_s |_{\it I}>$}

}
\startdata

VLIRG$_s$ & 19 &  0.047 $\pm$ 0.015&-20.48 $\pm$ 0.67 & -22.44 $\pm$ 0.45 & 1.96 $\pm$ 0.36 &  &  \\
VLIRG$_s$ (I-III)& 11 & 0.043 $\pm$ 0.015&-20.70 $\pm$ 0.72 & -22.55 $\pm$ 0.49 & 1.85 $\pm$ 0.39 &  &  \\
VLIRG$_s$ (IV-V)& 8 & 0.052 $\pm$ 0.012 &-20.25 $\pm$ 0.63 & -22.30 $\pm$ 0.31 & 2.05 $\pm$ 0.36 & & \\
VLIRG$_n$ & 29 & 0.047 $\pm$ 0.016\tablenotemark{b} &-17.85 $\pm$ 0.80 & -20.10 $\pm$ 0.45 & 2.25 $\pm$ 0.59 & 0.100$\pm$0.060 & 0.127$\pm$0.051 \\
VLIRG$_n$ (I-III)& 21 &  0.044 $\pm$ 0.016\tablenotemark{b}& -17.90 $\pm$ 0.87 & -20.06 $\pm$ 0.50 & 2.15 $\pm$ 0.62 & 0.100$\pm$0.069 & 0.119$\pm$0.055\\
VLIRG$_n$ (IV-V)& 8 & 0.052 $\pm$ 0.012 &-17.72 $\pm$ 0.55 & -20.22 $\pm$ 0.22 & 2.49 $\pm$ 0.41 & 0.100$\pm$0.022 & 0.151$\pm$0.026  \\
ULIRG$_n$ & 32 & 0.088 $\pm$ 0.044\tablenotemark{b} &-18.04 $\pm$ 1.68 & -20.11 $\pm$ 1.35 & 2.07 $\pm$ 0.69 &  &  \\
ULIRG$_n$ (I-III)& 20 & 0.104 $\pm$ 0.050\tablenotemark{b} &-17.59 $\pm$ 1.47 & -19.87 $\pm$ 1.26 & 2.28 $\pm$ 0.60 & & \\
ULIRG$_n$ (IV-V)& 12 & 0.082 $\pm$ 0.043 &-18.80 $\pm$ 1.80 & -20.50 $\pm$ 1.34 & 1.70 $\pm$ 0.65 & & \\
C-ULIRG$_s$ & 14 & 0.087 $\pm$ 0.042 &-20.58 $\pm$ 0.77 & -22.26 $\pm$ 0.57 & 1.68 $\pm$ 0.64 &  &  \\
C-ULIRG$_s$ (I-III) & 9 & 0.104 $\pm$ 0.042 &-20.50 $\pm$ 0.88 & -22.39 $\pm$ 0.45 & 1.89 $\pm$ 0.55 &  &  \\
C-ULIRG$_s$ (IV-V)& 5 & 0.057 $\pm$ 0.017 &-20.71 $\pm$ 0.58 & -22.01 $\pm$ 0.61 & 1.30 $\pm$ 0.66 & & \\
C-ULIRG$_n$ & 19 & 0.083 $\pm$ 0.037\tablenotemark{b} &-18.04 $\pm$ 1.20 & -20.21 $\pm$ 0.81 & 2.17 $\pm$ 0.66 & 0.128$\pm$0.103& 0.173$\pm$0.110 \\
C-ULIRG$_n$ (I-III) & 14 & 0.099 $\pm$ 0.040\tablenotemark{b} &-17.96 $\pm$ 1.34 & -20.26 $\pm$ 0.88 & 2.30 $\pm$ 0.61 & 0.128$\pm$0.108 & 0.170$\pm$0.122 \\
C-ULIRG$_n$ (IV-V) & 5 & 0.057 $\pm$ 0.017 &-18.25 $\pm$ 0.76 & -20.05 $\pm$ 0.36 & 1.80 $\pm$ 0.67 & 0.128$\pm$0.088 & 0.179$\pm$0.066 \\
W-ULIRG$_n$ & 13 & 0.094 $\pm$ 0.043\tablenotemark{b} &-17.90 $\pm$ 2.36 & -19.87 $\pm$ 1.96 & 1.97 $\pm$ 0.77 &  &  \\
W-ULIRG$_n (I-III)$ & 6 & 0.077 $\pm$ 0.03\tablenotemark{b} &-16.72 $\pm$ 1.36 & -18.95 $\pm$ 1.50 & 2.22 $\pm$ 0.59 &  &  \\
W-ULIRG$_n (IV-V)$ & 7 & 0.104 $\pm$ 0.06 &-19.20 $\pm$ 2.1 & -20.82 $\pm$ 1.65 & 1.62 $\pm$ 0.67 &  &  \\

\tablenotetext{a}{NOTES- Data for the ULIRG samples come from Surace et al. (1998, 2000). In the first column, subscripts s and n refer to 'systems' and 'nuclei' respectively. (I-III) and (I-V) refer to the interaction class according to the scheme by Surace (1998). See text.} 
\tablenotetext{b}{Counting only one nucleus per system}
\enddata
\end{deluxetable}

\clearpage




\figcaption{a. 'True' color images of the galaxies in the sample (see text). North is at the top, East at left. The horizontal line represents 5 kpc. }

\addtocounter{figure}{-1}
\figcaption{b. (continued)}

\figcaption{a. $B$,$V$, and $I$ images. North is at the top, East at left. The length of the arrow indicates 5$''$. The horizontal line represents 5 kpc.}

\addtocounter{figure}{-1}
\figcaption{b. Continued}

\addtocounter{figure}{-1}
\figcaption{c. Continued}

\addtocounter{figure}{-1}
\figcaption{d. Continued}

\addtocounter{figure}{-1}
\figcaption{e. Continued}

\addtocounter{figure}{-1}
\figcaption{f. Continued}

\addtocounter{figure}{-1}
\figcaption{g. Continued}

\figcaption{M$_B$-($B-V$) diagram for the objects in the VLIRG sample. The systems (full green dots) and nuclei (full red triangles) of the same object are connected by lines. We also represent temporal sequences generated with STARBURST99 (Leitherer 
et al. 1999) for an instantaneous burst of 10$^9 M_\odot$ (light-blue line), and a continuous starburst with a 
rate of 50$M_\odot$/yr (dashed sea-blue line). Squares indicate ages of 1, 2, 5, 10, 20, 50, and 100 Myr.  A Salpeter IMF, with masses between 0.1 and 120 $M_\odot$, and solar 
metallicity were assumed in both cases. The location of Mrk 1014 is also indicated with a cross (see text). }

\figcaption{M$_B$-$(B-V)$ and M$_I$-$(B-V)$ diagrams. The values corresponding to the VLIRGs are represented by full green dots (systems) and full red triangles (nuclei). The open triangles and  crosses correspond to the nuclei of the cool and warm ULIRGs, respectively, observed by Surace et al. (1998 and 2000). The open circles correspond to the cool-ULIRG systems. The shaded region represents the area of the knots (star clusters) detected by these authors with the {\it HST}.}

\figcaption{M$_B$-$(B-V)$ diagram for the nuclei and the systems in the VLIRG and ULIRG samples, where symbols distinguish among the different morphological classes as defined by Surace (1998) and Veilleux et al. (2002). Specifically,  full (magenta) squares represent {\it young} objects (classes I , II, and III), and open (blue) circles {\it old} objects (classes IV, and V). See text.}


\begin{thebibliography}{}

   \bibitem[2000]{AC 2000} Arribas, S., \& Colina, L.  2000, ApJ, 545, 228

   \bibitem[2000]{AC 2002} Arribas, S., \& Colina, L.  2002, ApJ, 573, 576
 
   \bibitem[2003]{B 2003} Beckwith, S.W., et al. 2003, Bulletin of the American Astronomical Society Meeting 202, 17.05
     
   \bibitem[2000]{Borne 2000} Borne, K., Bushouse, H., Lucas, R.A., \& Colina, L.  2000, ApJ, 529, L77
   
   \bibitem[2002]{Bushouse 2002} Bushouse, H. A., Borne, K. D., Colina, L., Lucas, R. A., Rowan-Robinson, M., Baker, A. C., Clements, D. L., Lawrence, A., \& Oliver, S. 2002, ApJS, 138, 1
   
   \bibitem[1989]{CCM 1989} Cardelli, J.A., Clayton, G.C., \& Mathis, J.S.  1989, ApJ, 345, 245 

   \bibitem[2003]{2003} Chapman, S.C., Blain, A.W., Ivison, R.J., \& Smail, R. 2003, Nature, 422, 695
   
   \bibitem[1989]{Clements 1996} Clements, D.L., Sutherland, W.J., Saunders, W., Efstathiou, McMahon, R.G., Maddox, S., Lawrence, A., \& Rowan-Robinson, M. 1996, M.N.R.A.S., 279, 459

   \bibitem[1989]{Clements 1996} Clements, D.L., Sutherland, W.J., McMahon, R.G., \& Saunders, W. 1996, M.N.R.A.S., 279, 477
   
   \bibitem[2001]{Colina 2001} Colina, L., Borne, K., Bushouse, H., Lucas, R. A., Rowan-Robinson, M., Lawrence, A., Clements, D., Baker, A., \& Oliver, S. 2001, ApJ, 563, 546
   
   \bibitem[2003] {Dickinson} Dickinson et al. 2003 in 'The Mass of Galaxies at Low and High Redshift'. Proceedings of the ESO Workshop held in Venice, Italy, 24-26 October 2001, p. 324
   
   \bibitem[2001]{Farrah 2001} Farrah, D., Rowan-Robinson, M., Oliver, S., Serjeant, S., Borne, K., Lawrence, A., Lucas, R. A., Bushouse, H., \& Colina, L., 2001, M.N.R.A.S., 326, 1333 
   
   \bibitem[1987]{Fisher} Fisher, K. B., Huchra, J. P., Strauss, M. A., Davis, M., Yahil, \& A., Schlegel, D. 1995, ApJS, 100, 69

   \bibitem[2003]{Frayer} Frayer, D.T., Reddy, N.A., Armus, L., Blain, A.W., Scoville, N.Z., \& Smail, I. 2003,  Astro-ph/0310656 (to be published in AJ, Feb04)
   
   \bibitem[2001]{Genzel} Genzel, R., Tacconi, L.J., Rigopoulou, D., Lutz, D., \& Tecza, M. 2001, ApJ, 563, 527 

   \bibitem[2003]{Giavalisco 2003} Giavalisco, M., et al. 2004, ApJ, 600, L93
   
   \bibitem[1998]{1998} Hauser, M.G., et al. 1998, ApJ, 508, 25

   \bibitem[1995]{Kim 1995} Kim, D.-C., Sanders, D.B., Veilleux, S., Mazzarella, J.M., \& Soifer, B.T. 1995, ApJS, 98, 129 

   \bibitem[2002]{Kim 2002} Kim, D.-C., Veilleux, S., \& Sanders, D.B. 2002, ApJS, 143, 277 
   
   \bibitem[1999]{1999} Lawrence, A., Rowan-Robinson, M, Leech, K., Jones, D.H.P., \& Wall, J.V. 1989, M.N.R.A.S., 240, 329
   
   \bibitem[1999]{1999} Lawrence, A., Walker, D., Rowan-Robinson, M., Leech, K.J., \& Penston, M.V. 1986, M.N.R.A.S., 219, 687 

   \bibitem [1999]{1999} Lawernce, A., Rowan-Robinson, M., Ellis, R. S., Frenk, C. S., Efstathiou, G., Kaiser, N., Saunders, W., Parry, I. R., Xiaoyang, X., \& Crawford, J. 1999 M.N.R.A.S., 308, 897

   \bibitem[1994]{1994} Leech, K.J., Rowan-Robinson, M., Lawrence, A., \& Hughes, J.D. 1994, M.N.R.A.S., 267, 253

   \bibitem[1999]{Leitherer} Leitherer, C., Schaerer, D., Goldader, J. D., Delgado Gonz\'alez, R.M.,  Robert, C., Kune, D. F., de Mello, D. F., Devost, D., \& Heckman, T. M. 1999, ApJS, 123, 3L

   \bibitem[1990]{1990} Melnick, J., \& Mirabel, I.F. 1990, Astron. Astrophys., 231, L19 

   \bibitem[1996]{1996} Murphy, T. W., Jr., Armus, L., Matthews, K., Soifer, B. T., Mazzarella, J. M., Shupe, D. L., Strauss, M. A., \& Neugebauer, G. 1996, AJ, 111(3), 1025
   
   \bibitem[1986]{1986} Rieke, G. H., \& Lebofsky, M. J. 1986, ApJ, 304, 326

   \bibitem[1996]{Sanders} Sanders, D.B., \& Mirabel, I. F. 1996, ARAA, 34, 749
   
   \bibitem[1990]{Saunders} Saunders, W., Rowan-Robinson, M., Lawrence, A., Efstathiou, G., Kaiser, N., Ellis, R. S., \& Frenk, C. S.. 1990, M.N.R.A.S., 242, 318

   \bibitem[2000]{Scoville} Scoville, N. Z., Evans, A. S., Thompson, R., Rieke, M., Hines, D. C., Low, F. J., Dinshaw, N., Surace, J. A., \& Armus, L. 2000, AJ, 119, 991
 
   \bibitem[1998]{Smail98} Smail, I., Ivison, R. J., Blain, A. W., \& Kneib, J.-P.  1998, ApJ, 507, L21
   
   \bibitem[1999]{Smail99} Smail, I., Ivison, R. J.; Kneib, J.-P., Cowie, L. L., Blain, A. W., Barger, A. J., Owen, F. N., \& Morrison, G. 1999, M.N.R.A.S., 308, 1061 
   
   \bibitem[1987]{Soifer} Soifer, B. T., Sanders, D. B., Madore, B. F., Neugebauer, G., Danielson, G. E., Elias, J. H., Lonsdale, Carol J., \& Rice, W. L. 1987, ApJ, 320, 238

   \bibitem[1988]{Stetson} Stetson, P.B., \& William, E. 1988, AJ, 96, 909
   
   \bibitem[1992]{Straus} Strauss, M. A., Huchra, J. P., Davis, M., Yahil, A., Fisher, K. B., \& Tonry, J., 1992 ApJS 83, 29
   
    \bibitem[1992]{Straus} Straus, M.A. et al. 1995, 'IRAS 1.2Jy redshift survey, Catalog 7185' (ref. from NED)

   \bibitem[2000]{Surace2000} Surace, J.A. 1998, Ph.D. thesis, Univ. Hawaii  
  
   \bibitem[2000]{Surace2000} Surace, J.A., Sanders, D.B., \& Evans. A.S. 2000, ApJ, 529, 170

   \bibitem[2000]{Surace1998} Surace, J.A., Sanders, D.B., Vacca, W.D., Veilleux, S., \& Mazzarella, J.M. 1998, ApJ, 492, 116

   \bibitem[1995]{Veilleux 1995} Veilleux, S., Kim, D.-C., Sanders, D.B., Mazzarella, J.M., \& Soifer, B.T. 1995, ApJS, 98, 171 

   \bibitem[2002]{Veilleux 2002} Veilleux, S., Kim, D.-C., \& Sanders, D.B. 2002, ApJS, 143, 315 
   
   \bibitem[1998]{Wu 1998a} Wu, H., Zou, Z. L., Xia, X. Y., \& Deng, Z. G. 1998a Astron. Astrophys. Suppl. Ser., 127, 521 

   \bibitem[1998]{Wu 1998b} Wu, H., Zou, Z. L., Xia, X. Y., \& Deng, Z. G. 1998b Astron. Astrophys. Suppl. Ser., 132, 182


\end{thebibliography}
\end{document}